\documentclass[10pt,preprint]{aastex}

\newcommand{\kms}{ km s$^{-1}$}

\newcommand\noh{N_\mathrm{OH}}
\newcommand\tex{T_\mathrm{ex}}
\newcommand\ut[2]{\,\textrm{#1}^{#2}}

\def\ddeg   {\hbox{$.\!\!^\circ$}}              % Degree over dot
\def\darcmin  {\hbox{$.\mkern-4mu^\prime$}}       % Arcminutes over dot
\def\dsec   {\hbox{$.\!\!^{\rm s}$}}            % Second over dot

\begin{document}

\title{GBT Observations of IC443: the Nature of OH(1720 MHz) Masers and OH Absorption} 
\author{J.W. Hewitt, F. Yusef-Zadeh} 
\affil{Department of Physics and Astronomy, Northwestern University, Evanston, IL 60208}

\author{M. Wardle} \affil{Department of Physics, Macquarie University, NSW 2109, Australia} 

\author{D.A. Roberts\altaffilmark{1}}
\affil{Department of Physics and Astronomy, Northwestern University,
Evanston, IL. 60208}
\altaffiltext{1}{Astronomer, Adler Planetarium \& Astronomy Museum, 1300 S. Lake Shore Drive, Chicago, IL. 60605}
\and
\author{N.E. Kassim} \affil{Code 7213, Remote Sensing Division, Naval Research Laboratory, Washington, DC 20375}

\begin{abstract}
We present spectral line observations of the ground state transitions of hydroxyl(OH) toward supernova remnant(SNR) IC443 carried out with the Green Bank Telescope. Weak, extended OH(1720 MHz) maser emission with OH(1667,1665,1612 MHz) absorption is detected along the southern extent of the remnant where no bright compact maser sources have been previously observed. These newly detected SNR-type masers are coincident with well-known molecular clumps and a ridge of shocked H$_{2}$ emission indicative of the SNR shock front interacting with the adjacent molecular cloud. Simultaneous observation of all four ground-state transitions of OH permits us to fit physical conditions of the shocked gas at the interaction site. A simple two-component model for the line profiles yields the physical parameters for detected regions of maser emission including excitation temperature, OH column density and filling factor. Observed line profiles suggest the shock is largely propagating toward the line-of-sight in the region of these newly identified weak masers. The implications of shock geometry and physical parameters in producing spatially extended OH maser emission in SNRs are explored. We also present VLA radio continuum observations at 330 MHz for comparison with OH line observations of the remnant.
\end{abstract}

\keywords{ISM: general --- shock waves --- supernova remnants: individual(\objectname{IC443}) --- {\it Facilities:} \facility{GBT}, \facility{VLA}}

\section{Introduction}
The supernova remnant(SNR) IC443 (G189.2+3.0) is one of the best sources in which to study the physical processes of a supernova remnant interacting with adjacent molecular clouds. Located at a distance of 1.5 kpc in a direction largely free of the confusion present toward the Galactic plane, it was well suited for some of the earliest studies of molecular tracers of shock excitation (DeNoyer 1979; Huang, Dickman \& Snell 1986; Burton 1987; Tauber et al. 1994). Evidence of IC443's interaction was first observed as large line-width, negative velocity neutral hydrogen(HI) by DeNoyer (1978). Subsequent carbon monoxide(CO) observations showed a slow shock had been induced across a ridge of molecular material in the southern half of the remnant (DeNoyer 1979). Notably, the ratio of OH/CO column densities was observed to be enhanced by two orders of magnitude compared with preshock cloud. Huang, Dickman \& Snell (1986) identified a series of shocked CO clumps (designated A-H) along this southern ridge, with sizes ranging from 0.5 to 5 pc. Further high-resolution study revealed shocked molecular line emission tracing out a ring of perturbed molecular gas about the center of the remnant (Dickman et al. 1992; van Dishoeck, Jansen \& Phillips 1993).

%Quite a different interaction picture is suggested by striking optical filaments along the northern extent of IC443. Optical spectroscopy of these filaments indicates a shock velocity of V$_S$ $\approx$ 65 to 100\kms, sufficient to partially destroy interstellar grains (Fesen \& Kirshner 1980). It is clear that IC443 is interacting with two very different media and that no simple global shock model can be invoked. This transition between the northern J-type  and southern C-type shocks is beautifully traced out by the remnants near-IR colors (Wang \& Scoville 1992; Rho et al. 2001).

As a rich, well-studied laboratory for shock chemistry, IC443 was again a prominent target for the study of OH following the realization by Frail, Goss \& Slysh (1994) that maser emission from the 1720.530 MHz transition is a signpost of interaction between SNRs and molecular clouds. Theoretical considerations yield strict constraints on the conditions necessary for SNR OH(1720 MHz) masers: moderate temperatures of 50-125 K, densities of 10$^5$ cm$^{-3}$, and OH column densities of the order 10$^{16}$-10$^{17}$ cm$^{-2}$ (Lockett, Gauthier \& Elitzur 1999). Masers are thought to be collisionally excited when a non-dissociative shock propagates into a sufficiently dense molecular cloud.

While these conditions are satisfied throughout much of the shocked ring of molecular material present in IC443, only a single molecular clump G near the projected center of the remnant hosts identified masers (Frail et al. 1996). Subsequent high-resolution study has used this species of maser as a unique probe of post-shock magnetic field, as well as density and temperature at arcsecond resolution (Claussen et al. 1997). Hoffman et al. (2003) has resolved the brightest compact masers in clump G, revealing 60 AU cores surrounded by diffuse halos extending up to 300 AU in size. While the entire shocked clump appears capable of producing maser emission only a small fraction does ($\sim$10$\%$). However, such studies of IC443 remain limited to clump G.

The lack of observed compact OH(1720 MHz) emission toward the southern ridge of IC443 underscores the importance of geometry in the amplification of OH(1720 MHz) emission. Claussen et al. (1997) suggests that only when the shock is transverse to the line of sight can OH(1720 MHz) masers be observed; it is in this geometry that maximal velocity coherence and pathlength occur. As the shock propagates into the adjacent cloud, water molecules produced by the shock are dissociated into OH indirectly by thermal X-rays emitted from hot gas in the SNR interior (Wardle 1999). This OH enhancement (OH/H$_2$$\ga$10$^{-6}$) behind the shock can be detected as thermal absorption against the bright background continuum of the remnant, regardless of shock geometry. Absorption profiles show the velocity dispersion of the gas toward the line of sight, which permits shock geometry traced by OH to be determined.

Motivated by this unique opportunity to probe very specific conditions just behind the shock front we have observed the four ground state transitions of OH over the entire extent of IC443. Sections 3.1 and 3.2 present our detections of OH(1720 MHz) maser emission and ground-state OH absorption, respectively, across the southern portion of the remnant. Simple modeling of the identified maser sources is given in $\S$3.3. Kinematics of the masing regions and and the nature of extended maser emission are discussed in $\S$4 in relation to the global shock interaction picture. Our 330 MHz radio continuum image of IC443 is used to place the shock interaction picture presented by line observations in context with the radio morphology of the remnant.

\section{Observations}
\subsection{Spectral Line Data}
The Green Bank Telescope (GBT) of the National Radio Astronomy Observatory\footnote{The National Radio Astronomy Observatories is a facility of the National Science Foundation, operated under a cooperative agreement by Associated Universities, Inc.} was used on 2003 August 15 and October 7 to observe the four ground-state transitions of OH at 1612.231, 1665.4018, 1667.359 and 1720.530 MHz in left- and right-circular polarizations. The observations presented here are part of a larger observing program which mapped all four ground-state transitions of OH for 21 remnants. A more detailed account of the results of this survey will be given in a future publication.

A map of the entirety of IC443 was constructed using a grid of 210 pointings which extends 51$'\times$48$'$ about the projected center $\alpha$,$\delta$($J2000$) = 6$^{\rm h}$16$^{\rm m}$56$^{\rm s}$, +22$^{\rm o}$29$\arcmin$8$\arcsec$. The GBT has a beam-size of 7$\darcmin$2 at 1.7 GHz, so pointings are spaced by 3$\darcmin$3 to correspond to the maximal spatial resolution for Nyquist sampling. Each pointed spectra has a RMS noise level of 15-20 mJy channel$^{-1}$ with a velocity resolution of 0.28\kms\ over the 12.5 MHz bandwidth. Calibration via in-band frequency switching limits the effective velocity coverage to 400\kms\ (1.17 MHz) about the rest frequency of each line.

For modeling it was necessary to obtain the continuum level from our observations. The difference of the system temperature measured on source compared to our blank sky position ($\Delta$T$_{sys}$) is used to derive the continuum temperature (T$_{src}$) using the equation:
\begin{equation}
\Delta T_{sys} = 3 \Delta A + \eta_B  T_{src}  {\rm exp}(- \tau A_{src})
\label{eq:tsys}
\end{equation}
where $\eta_B$ is beam efficiency (0.94), $\tau$ is atmospheric opacity (0.011 at L-band), and $\Delta$A = 1/sin($\Theta_{src}$)--1/sin($\Theta_{sky}$) where $\Theta$ is the elevation of observations on source and on blank sky. We estimate the measured continuum temperature to be accurate within 10$\%$.
Calibration, image processing and profile fitting utilized the AIPS++\footnote{AIPS and AIPS++ are Astronomical Image Processing Systems developed by the NRAO} and MIRIAD\footnote{MIRIAD is the Mutlichannel Image Reconstruction, Image Analysis and Display system of the Berkeley-Illinois-Maryland Association (BIMA)} astronomical processing systems.

\subsection{Continuum Data}

The 330 MHz radio continuum observations of IC443 were obtained as part of an unrelated project to survey $\gamma$-ray sources from the EGRET catalog (Hartman et al. 1999 and references therein). One of these sources, 2EG J0168+2234, lies close enough to IC443 with respect to the VLA's primary beam ($\sim$2$\ddeg$5 at 330 MHz) so as to provide a good serendipitous image. The observations were obtained in March and July of 1996 using the VLA's C and D configurations, respectively. The data were acquired in 2 IFs centered at 327.5 and 333 MHz in each of two circular polarizations. Since the observations were part of a survey, the on-source integration time was limited to approximately 40 and 30 minutes in the C and D configurations, respectively. While limited, these cycling-snapshot multi-configuration observations nonetheless provided a sufficient range of spatial frequencies to recover most of the structure in this large ($\sim$45$'$) object.

Post-processing of low-frequency VLA data uses procedures similar to those at higher frequencies although certain details differ. Phase calibration serves to compensate for ionospheric phase effects but because the ionosphere is fairly coherent across the VLA in these configurations single observations of 0137+331 (3C48) and 0542+498 (3C147) were sufficient for both flux density, bandpass and initial phase calibration. In order to combat RFI, the data were acquired in multi-channel continuum mode (32 channels per polarization) at higher spectral resolution than needed for imaging. Furthermore, the large field of view requires full three-dimensional deconvolution to remove the effects of the VLA's non-coplanar baselines. To achieve this, the data were edited and calibrated in standard AIPS$^{2}$, while NRAO's Software Development Environment polyhedron algorithm "DRAGON" (Cornwell \& Perley 1992) was utilized for performing successive loops of self-calibration and imaging. Comparison of the position of NVSS sources detected in our image with their positions in the NVSS catalog (the latter good to $\sim$1$\arcsec$) showed that the astrometry of the radio map is accurate to at least a few arc-seconds.

The final 330 MHz radio continuum image, uncorrected for primary beam attenuation, is shown in Figure \ref{fig:continuum}. The synthesized beam is 74$\arcsec\times$64$\arcsec$ at a position angle of 48 degrees, and the rms noise is 6 mJy beam$^{-1}$. After removal of unrelated point sources, the peak brightness of the SNR is 870 mJy beam$^{-1}$ and the total integrated flux is 182 Jy after primary beam correction. Because cycling snapshots are notoriously limited in their ability to recover emission from such an extended object these flux measurements should be taken as a lower limit. Comparison with single dish measurements of this object compiled by Erickson \&\ Mahoney (1985) give flux densities well over 200 Jy, even at somewhat higher frequencies. However, our 330 MHz continuum image does reproduce most of the structure revealed in the deeper 327 MHz WSRT map (Braun \&\ Strom 1986) and is more than adequate to provide a useful finder atlas for referencing our higher frequency OH observations.

\section{Results}
\subsection{OH(1720 MHz) Maser Emission}
We have identified three sites of OH(1720 MHz) emission in IC443 with $\sim$1\kms\ line-widths indicative of stimulated emission. At the 7$\darcmin$2 resolution of these observations we identify a bright (S$_p$=3.6 Jy) maser feature associated with molecular clump G, and two spatially extended maser features of S$_p$$\approx$0.15 Jy which are coincident with molecular clumps B and D. Table \ref{tbl:gaussfits} gives the Gaussian fits to these emission features. Here our maser designations follow those of molecular clumps presented in Huang, Dickman \& Snell (1986, see Figure  1).

The bright maser G has been previously detected with 1.5$\arcsec$ resolution VLA observations as six compact maser spots all with velocities near --4.5\kms\ (Claussen et al. 1997). All six compact masers fall well within one GBT pointing. The single dish measured flux density of 3.64 Jy is consistent with the VLA measured flux density of 3.89 Jy, accounting for single-dish flux calibration errors of 10\% . The implication is that all maser emission from this clump is present on the very small scales resolved by interferometric observations. Right circular polarization is detected at a level of 13$\%$, consistent with the earlier results of Claussen et al. (1997) and Hoffman et al. (2003).

For the first time OH(1720 MHz) maser emission is detected along the southern extent of IC443 at a level 20 times weaker than the bright compact maser G. Clumps B and D are observed with a peak brightness of 158 and 145 mJy and at a velocity of --6.14 and --6.85\kms, respectively. Claussen et al. (1997) were not able to detect weak maser features toward these clumps down to 5 mJy beam$^{-1}$ channel$^{-1}$ with high resolution VLA observations. Maps of the velocity integrated line flux density (moment zero) in Figure  2 show the newly detected emission features extending several arcminutes along the southern ridge of IC443 following the nonthermal radio continuum at 330 MHz. Circular polarization was not detected for these weak, extended features.

We consider two possible explanations for the weak 1720 MHz emission along the southern ridge. First, the observed emission could be due to multiple compact masers coincident with clumps B and D. However, previous VLA observations would have been sufficient to detect these sources if the masing regions are smaller than ~1$'$ in size, as was observed toward clump G. Therefore we suggest that these newly identified masing regions are spatially extended, having properties similar to the weak, extended regions of maser emission observed in remnants W28, 3C 391, G359.1--0.5, G357.7+0.3 and G357.7--0.1 (Yusef-Zadeh, Uchida \& Roberts 1995; Yusef-Zadeh et al. 1999).
The narrow line-widths observed ($\la$2\kms) are indicative of non-thermal maser emission. Brightness temperatures for these extended masers are $\sim$2 K, uncharacteristically low for compact masers (T$_B\approx$10$^4$ K) but consistent with the several arcminute scales and lower brightness temperatures (T$_B\approx$10-2500 K) of extended maser emission identified by Yusef-Zadeh et al. (1999). The implications of extended maser emission are discussed in $\S$4.2.

It is unlikely that the newly identified maser emission is a result of time-variable masing that has brightened significantly since VLA observations were conducted. Goss (1968) first detected OH(1720 MHz) maser emission in W28 and no time-variability has yet been observed for that source (Frail et al. 1994). We observe no change in the brightness of emission from clump G compared to previous observations over an 8 year period (Hoffman et al. 2003; Claussen et al. 1997).

\subsection{Thermal OH(1667,1665,1612 MHz) Absorption}

Simultaneous observations of all four transitions of OH confirm that masing occurs only at 1720 MHz and traces the kinematic structure against the nonthermal continuum of the remnant. The strongest absorption and clearest velocity features are observed in the 1667 MHz transition. Figure 3 shows line profiles of the four observed transitions toward maser clumps G, B and D. %Multi-component Gaussian fits to the absorption profiles observed for clumps G, B and D are given in Table 2. 
The absorption is found strongest in the 1667 MHz transition, and weakest at 
1612 MHz. No significant absorption is detected in the 1720 MHz transition.

The absorption profile toward clump G can be well-fit by two Gaussian components: (1) a stronger, narrow line with $\Delta$V$\approx$5\kms\  at V$_{LSR}$$\approx$--3\kms; and (2) a weaker, wide line with $\Delta$V$\approx$20\kms\ at V$_{LSR}$$\approx$--7 \kms. This profile is characteristic of transverse acceleration, where the absorption is strongly peaked near the systemic velocity of the remnant, and is somewhat broadened by thermal excitation of OH gas from the passage of the shock (Figure  3a). Furthermore, there is no clear delineation in velocity between pre- and post-shock absorption, though close inspection of the spectra of clump G reveals a slight asymmetrical broadening toward negative velocities indicating part of the shock is inclined slightly toward the line of sight. Other molecular species observed toward clump G show the same characteristic profile (Burton 1987; Dickman et al. 1992; van Dishoeck, Jansen \& Phillips 1993; Tauber et al. 1994). A transversely propagating shock is also needed to maximize the coherent path-length and amplification of the observed maser \cite{claussen97}.

Absorption profiles throughout the southern extent of the remnant show broad, asymmetrically blue-shifted line-widths of 30-40\kms. Figures 3b and 3c show line profiles toward masers B and D, respectively.  Strongly asymmetric blue-shifted broadening of the post-shock absorption profile suggests that the shock is accelerating the molecular gas toward the observer. In this geometry gas in all phases of shock acceleration is  present along the line of sight causing this spectral delineation of velocity components relative to the systemic velocity of the remnant. A narrow line at V$_{LSR}$ $\approx$ --3\kms with a width $\Delta$V $<$ 5\kms delineates the ambient unshocked gas.
The shocked gas along the southern ridge is fit by two Guassian components: (1) a strong, broad line with $\Delta$V $\approx$ 10\kms at V$_{LSR}$ $\approx$ --10 to --12\kms; and (2) a shallow, very broad line with $\Delta$V $\approx$ 20 to 30\kms\ at  V$_{LSR}$ $\approx$ --15 to  --25\kms. This decomposition of the asymmetric broadening into two Gaussian components can be non-unique, but is offered here as a means of first-order insight into shock dynamics.

Thus, the shock geometry is inferred to be transverse to the line of sight through clump G, and toward the line of sight along the southern ridge that includes clumps B and D, where extended masers are detected. The association between shock geometry and extended maser kinematics is discussed further in $\S$4.1. 

There is good agreement between the profiles of OH and other molecular shock tracers, particularly CO, observed at comparable resolution. Snell, et al. (2005) observed H$_2$O, CI, $^{13}$CO(5-4), $^{12}$CO(1-0), and HCO$^{+}$ for clumps B and G (see Figures 1 and 3, respectively in that work). $^{13}$CO(5-4), thought to be created by the shock, mirrors the kinematic picture shown by OH absorption in clumps B and G. However, for both clumps H$_2$O profiles extend beyond the largest velocities at which we detect OH absorption: in clump G H$_2$O emission is is observed out to --30 \kms and in clump D out to velocities of --80 \kms . Snell et al. (2005) have suggested both a fast J-type and slow C-type shock are needed to explain the observed shocked molecular species along the southern extent of IC443. The observed broader velocity components only for H$_2$O is consistent with some H$_2$O being liberated behind a J-type shock, while OH is mainly created behind the C-type shock front. It is interesting to note that given the observed OH column densities (Table 2) the column of H$_2$O is orders of magnitude less than predicted by theory. This apparent lack of H$_2$O is not accounted for in any present theory for the production of OH(1720 MHz) masers.

\subsection{Emission and Absorption models}

In this section we estimate the parameters of the gas responsible for producing the observed line profiles using a simple two-component model. As shown in Figure 4, our model fitting includes a maser component and a component to fit the broad absorption arising from the shocked gas. We do not include ambient gas absorption in these simple models. Each component is characterized by its line-of-sight OH column density $\noh$, kinetic temperature $T_k$, mean and FWHM velocities, $v_0$ and $\Delta v$, and a beam filling factor $f$. The H$_2$ density is fixed at $10^5$\,cm$^{-3}$, typical of what is expected in OH(1720 MHz) maser-emitting regions (Lockett, Gauthier \& Elitzur 1999). Excitation by collisions and partial trapping of photons then determines the excitation temperature $\tex$ of the 4 transitions and the line-center optical depth. Model fits are done independently of the Gaussian fits used for the discussion of kinematics in earlier sections. 

The observed intensity for a particular line is
\begin{equation}
     I_\nu=(1-f)I_{0\nu}+f\left[I_{0\nu}\exp(-\tau_\nu)+
     B_\nu(\tex)(1-\exp(-\tau_\nu))\right]
     \label{eq:Inu}
\end{equation}
where $I_{0\nu}$ is the background continuum intensity and $B_\nu$ is
the Planck function. Here the obscuration of one component by another 
is neglected. Assuming that $|\tex|\gg 0.08$\,K, the
Rayleigh-Jeans approximation holds, and eq. (\ref{eq:Inu}) can be recast as
\begin{equation}
     T(v) - T_0 = f (\tex - T_0 )  (1 - \exp(-\tau(v)))
     \label{eq:T_v}
\end{equation}
We write the optical depth as
\begin{equation}
     \tau(v) = \tau_0 \exp\left(-\frac{(v-v_0)^2}{2\sigma^2}\right)
     \label{eq:tau_v}
\end{equation}
where $\tau_0$ is the optical depth at line center and $\sigma = \Delta
v / (8 \ln 2)^{1/2}$ and we have adopted a Gaussian line profile.

The optical depth at line center is
\begin{equation}
	\tau_0 = \left(\frac{\ln 2 }{\pi}\right)^{1/2}
	\frac{\lambda^3 A}{4\pi} \frac{\noh}{\Delta v}
	\frac{g_2 x_1 }{g_1}
	\left[1 - \exp\left(-\frac{\Delta E}{k\tex}\right) \right]
     \label{eq:tau0}
\end{equation}
where $g_1,x_1$ and $g_2,x_2$ are the degeneracies and fractional
populations of the lower and upper states, and $\lambda, A$
and $\Delta E$ are the wavelength, Einstein A value, and energy difference for
the transition.

For the four levels in the rotational ground state of OH, usually
$|T_{\mathrm{ex}}| \gg \Delta E/k \sim 0.08$\,K; in addition the higher
rotational states are not significantly populated.  This means that
the term in the square brackets in eq.\ (\ref{eq:tau0}) can be
approximated as $\Delta E/k\tex$ and also that the four levels in
the ground rotational state are very nearly populated according to their
degeneracies (ie. $x_{1,2} \approx g_{1,2}/16$).  Thus
\begin{equation}
	\tau_0 = a\,
	\frac{\noh(10^{15}\ut{cm}{-2})}{T_{\mathrm{ex}}(\textrm{K}) \Delta
	v(\textrm{km/s})}
	\label{eq:tauOH}
\end{equation}
where the constant $a$ is 0.4540, 2.3452, 4.2261, and 0.4846 for the
1612, 1665, 1667 and 1720\,MHz lines respectively.  Note that $\tex$
and $\tau_0$ are negative for an inverted transition.

Eqs (\ref{eq:T_v}), (\ref{eq:tau_v}) and (\ref{eq:tauOH}) form the
basis of a simple model for the observed lines, with parameters $\noh$,
$T_k$, $f$, $v_0$, $\Delta v$.  The excitation temperatures are
determined by an LVG model of the collisional excitation of the OH
molecules within a clump, including the effect of line overlap and
radiative transfer in the maser line, following
Lockett, Gauthier \& Elitzur (1999).  Given the uncertainty inherent in the
single-dish data, we do not perform a detailed fit to the line
profiles (which in any case are not Gaussian); instead we have
adjusted parameters until a reasonable match is found.  The derived
parameters should therefore be regarded as illustrative rather than
definitive.

To match the observed profiles two Gaussian components are required: a narrow component to match bright 1720 MHz maser emission, plus a broad component responsible for the bulk of absorption in the other lines. Fits of our simple models for clumps B, D and G are presented in Figures 4-a,b,c respectively. The net contribution of components is shown as a bold dark line against the observed profile. The broad component has a peak optical depth of order unity for Clump G and appears optically thin for Clumps B and D (see Table 2). For the line of sight to clumps B and D a single broad component with parameters very similar to the broad component for the clump G model, except with a significantly lower column density, successfully reproduces the line profiles in all four transitions. Given the observed OH(1720 MHz) narrow emission component, the column densities derived from our simple models are as expected from models of OH production in shock waves (Wardle 1999) and collisional excitation of OH masers (Lockett, Gauthier \& Elitzur 1999).

For the narrow component the main lines do not appear in the expected rough 5:9 ratio for optically thin absorption with similar excitation temperatures. We observe $\tex \ll T_0\approx 30\,K$, and the brightness temperature at line center only drops by 0.3 K. The relatively bright and narrow 1720 MHz maser component requires a much smaller filling factor to avoid introducing sharp absorption feature into the absorption lines. We therefore conclude the line is optically thick with a beam filling factor $f \sim$ 0.01 for this component. We note that some confusion between pre- and post-shock gas may be unavoidable for modeling of clump G given the overlapping velocities of ambient cloud and post-shock gas. For clumps B and D the separation between pre- and post-shock velocity components is sufficient to avoid significant confusion between the two. This is apparent from the significant absorption present at --3 \kms which extends below the dark line indicating the simple model fit in Figures 3b and c.

\subsection{OH Abundances}
To further demonstrate that OH is enhanced behind the passage of the SNR shock, as predicted by theory we can compare pre- and post-shock columns and abundances. The column density of OH can be estimated using 
\begin{equation}
        N_{\rm OH} = 2.2785 \times 10^{14} \quad \tex \int \tau_v \quad dv \rm{\quad cm^{-2}}
        \label{eq:NOH}
\end{equation}
for the 1667 MHz line (Crutcher 1977). For clump B, in which the seperation between pre- and post-shock components is clearest, we find N$_{\rm OH}$ = 1.2$\times$10$^{13}$ $\tex$ cm$^{-2}$ for the pre-shock gas. Here we have used the Gaussian fit to the pre-shock component presented in $\S$3.2 to derive $\tau_{peak}$=0.026 and $\Delta$V=2\kms. From our simple modeling of the broad post-shock gas we find N$_{\rm OH}$ = 1.1$\times$10$^{16}$ cm$^{-2}$ and $\tex$ = 7.5 K at 1667 MHz. Assuming the excitation temperature for the pre-shock gas is no more than that found for the shocked gas, we find an enhancement in OH column greater than two orders of magnitude. This enhancement in column density is indicative of the production of OH behind the shock front given that the shock through clump B is propogating mostly along the line-of-sight and both the pre-shock and broad component post-shock gas are thought to fill the beam.

To calculate OH abundances we used characteristics of the pre- and post-shock gas derived for clump B by van Dishoeck, Jansen \& Phillips (1993). The pre-shock column of H$_2$ is found to be 2$\times$10$^{21}$ cm$^{-2}$ yeilding OH/H$_2$=4.5$\times$10$^{-8}$. For the post-shock gas the CO column density is found to be 1.1$\times$10$^{17}$ cm$^{-2}$ for a gas with T=80 K and n$_{H_{2}}$=10$^5$ cm$^{-2}$ and $\Delta$V = 20\kms, comparable to the broad component we have modeled for clump B. Assuming CO/H$_2$ is 10$^{-4}$, typical of interstellar clouds, the post-shock abundance of OH is 6$\times$10$^{-6}$ indicative of an OH enhancement exceeding two orders of magnitude. We note that molecular clumps in IC443 are typically $\sim$1$\arcmin$ in extent (van Dishoeck, Jansen \& Phillips 1993), which is significantly smaller than the 7$\darcmin$2 beam of these observations. Beam dilution may be significant, up to a factor of 50, however we argue that such a large dilution value is unlikely. Large-scale CO observations show the pre-shock gas is extended over regions greater than the GBT beam (Seta et al. 1998). Our simple modeling of the post-shock gas finds a filling factor of order unity for the broad component of OH absorption, consistent with multiple shocked clumps of arcminute size present within a single pointing. Even accounting for the uncertain effects of beam dilution, it is clear that OH is enhanced by the passage of the SNR shock.

\section{Discussion}

\subsection{Kinematics}
OH(1720 MHz) maser amplification favors a very specific geometry for the SNR-MC
interaction in which the shock propagates across the sky, transverse to the 
line of sight. In this geometry the velocity gradient of the masing gas is 
minimized and the largest coherent path lengths are achieved \cite{frail96}. 
Thus the observed line-of-sight velocities of SNR-type masers are commonly 
found within a few\kms\ of the systemic velocity even when distributed
throughout the remnant.

The masers associated with IC443 all lie within 2\kms\ of the systemic velocity (--5\kms), and the compact maser emission of clump G has been associated with a shock propagating transverse to the line of sight as is expected \cite{claussen97}. OH absorption profiles toward clumps B and D show a clear delineation between pre- and post-shock velocity components with asymmetrically blue-shifted shock broadening, indicative of shock acceleration toward the line of sight. This face-on geometry is observed along the entirety of the southern ridge of IC443 tracing the shock interface with adjacent molecular material. In this face-on geometry the OH pathlength and maser intensity are greatly reduced, and masing gas is present at distinctly different velocity than the bulk of the gas observed in absorption.

To reconcile these disparate kinematic results, it is necessary to  consider the global kinematic structure revealed by high resolution molecular line observations. Dickman et al. (1992) mapped the entirety of IC443 tracing shocked gas in the J=1-0 lines of CO and HCO$^+$. The observed shocked clumps are found in a ring $\sim$9 pc across, tilted by $\sim$50$^o$ from the line of sight. Observations of $^{12}$CO 3-2 transition by van Dishoeck, Jansen \& Phillips (1993) resolve molecular clumps A-H into smaller sub-clumps embedded in a sinuous ring of molecular material.  At 1.6 GHz the GBT beam samples regions with 7$'$ resolution, and may include multiple molecular sub-clumps within one pointing. 

For all  positions in which 1720 MHz maser emission is observed, comparison with CO 3-2 spectra indicates the existence of transverse shocks in which an optimal geometry for OH(1720 MHz) maser emission can arise. For a comparison of CO 3-2 with our OH observations we note that Figure 3 of van Dishoeck, Jansen \& Phillips (1993) presents spectra of CO sub-clumps R10-13 contained within clump B, R17-19 within clump D, and R1-2 within clump G. For masers B and D, while the brightest sub-clumps show highly blue-shifted profiles, weaker sub-clumps are also identified within each GBT pointing that show CO 3-2 profiles indicative of transversely directed shocks. At yet higher resolution, CO 3-2 sub-clumps within clump B show drastic variations in observed line profiles on scales less than 1$\arcmin$ (Figure 5 of van Dishoeck, Jansen \& Phillips 1993). Small scale variations in shock stucture are consistent with profile modeling in $\S$3.3. Both a broad and narrow component are needed to explain the observed profiles, representing the face-on and transversely accelerated clumps, respectively. Modeling of the masing gas also indicates OH column densities are a factor of 5 to 20 lower for masers B and D than for maser G. Maser emission arises in the low-column density, transversely shocked clumps throughout the southern ridge.

Globally, blue-shifted OH absorption profiles are never found to exceed $\approx$ --20 to --30\kms. This is consistent with the ring of molecular material undergoing a non-dissociative C-type shock. As shown by Lockett, Gauthier \& Elitzur (1999) and Wardle (1999), only C-type shocks are capable of producing the conditions necessary for the production of OH(1720 MHz) masers. This is in contrast with observed o-H$_2$O profiles which extend out to --80\kms for clump B. It has been suggested that a faster J-type shock is needed to explain these large velocity offsets for shocked molecularl material and the observed [C II] emission (van Dishoeck, Jansen \& Phillips 1993; Snell et al. 2005). However, OH appears to only be present at lower velocities consistent with a slow shock.

\subsection{Extended Maser Emission}
As discussed in $\S$3.1, the newly detected masers B and D are thought to 
be spatially extended regions of masing gas. In previous cases compact maser 
sources have been identified with surrounding extended halos of maser-like 
emission at significantly lower brightness temperatures: typically 
T$_{B}$$\approx$1-2500 K whereas compact maser sources are typically 
T$_{B}$$\approx$5$\times$10$^4$ K \cite{fyz99}.
Such a low brightness level and a large spatial extent would have made previous
interferometric observations with the VLA insensitive to such extended maser
emission. 

Gaussian fits to these newly detected features show spatial extension over a few GBT beam widths. These large scale masers suggest a low-gain population inversion of OH is created over extended regions of the ambient cloud by the passage of the shock front. This is supported by the correlation between regions of extended maser emission and shocked H$_2$ emission indicative of the SNR shock front interacting with the adjacent molecular cloud. Such is the case for the newly identified extended maser emission in IC443, which is present along the southern ridge of H$_2$ emission \cite{burton87} As discussed in $\S$4.1, the molecular environment at these interaction sites is clumpy on scales smaller than the resolution of our observations \cite{tauber94}, so further observations which can resolve the weak, extended regions of masing are needed to probe the shock-cloud interaction on scales comparable to the clumpiness of the medium.

The newly detected extended masers in IC443 are the first such instance of weak, extended maser emission with no associated compact masers. A velocity gradient of a few\kms\ is present over the expanse of the extended maser emission suggesting that for large-scale masers acceleration is unlikely to be entirely perpendicular to the line of sight, as is thought to be the case for compact masers (Lockett, Gauthier \& Elitzur 1999). IC443 may not be unique in this respect; extended masers in 3C 391 appear at significantly different velocities from compact masers. Further modeling of maser amplification under different shock geometries is necessary to better understand the stimulation of these extended masers.

\section{Conclusions}
Sensitive observations of the four ground-state transitions of OH
over the entirety of IC443 lead to the following conclusions.

1. Thermal OH absorption is detectable against the background continuum
throughout the southern extent of IC443, and reveals the direction of shock 
propagation. The western extent (IC443G) is undergoing transverse motions
while bulk motions throughout the southern extent (IC443B,D) indicate a line 
of sight acceleration.

2. Extended maser emission is detected along the southern ridge of IC443 toward molecular clumps B and D with low brightness temperatures and velocities systemic to the remnant and associated ambient medium. The disparity between observed maser emission and absorption profiles is indicative of cloud clumpiness and kinematic variations on scales less than 7$\darcmin$2, the resolution of these observations.

3. The observed OH line profiles cannot be fit with a single shock component. 
Two component modeling derives $v_0$, $\Delta$$v$, N$_{OH}$, T$_k$, $f$
and T$_{ex}$ for detected regions of maser emission, consistent with
theoretical constraints for OH(1720 MHz) masers in supernova remnants.

Further investigation of weak extended masing must be conducted at
higher resolutions to constrain the spatial scales and shock geometries in
which extended maser emission occurs.

\acknowledgments{We are very grateful to Ron Maddalena and Jim Braatz for their assistance with observations at Green Bank and helpful discussions. Support for this work was provided by the NSF through awards AST-0307423, and GSSP 2-0005 from the NRAO. Basic research in radio astronomy at the Naval Research Laboratory is supported by the Office of Naval Research.}

\clearpage

%table 1, emission fits
\begin{table}[h]
\centering
\caption{OH(1720 MHz) Emission Gaussian Fits}
\vspace{10pt}
\begin{tabular}{crrcccccc}
\hline
Clump & $\alpha_{J2000}$ & $\delta_{J2000}$ &  S$_P$  &  V$_{LSR}$  &  $\Delta$$V$  &  $\theta_{MAJ}$  &  $\theta_{MIN}$  & P.A.  \\
 & h m s & $\degr$ $\arcmin$ $\arcsec$ & mJy & \kms\ & \kms\ & $\arcmin$ & $\arcmin$ & $\degr$ \\
\hline
B & 06 16 40.3 & +22 23 06 &  158(22) & --6.14(0.08) & 1.04(0.09) & 13.8 & 6.5 & 77.4(0.1)\\
D & 06 17 53.2 & +22 23 50 &  145(17) & --6.85(0.07) & 1.76(0.11) & 11.4 & 6.5 & 46.8(0.1)\\
G & 06 16 47.0 & +22 32 01 & 3641(35) & --4.55(0.04) & 0.84(0.05) &  7.7 & 6.9 & 57.2(0.4)\\
%? & 06 17 27.9 & +22 21 33 &  138(17) & -1.39(0.06) & 0.79(0.14) & 13.8 & 6.5 & 77.4(0.1) \\
%the above line is the weird new maser feature...
%G & 06 16 45.4  & +22 32 16  & 5112(56) &  -4.48(0.01)  &  0.82(0.02)  & 9.3  &  8.7  &  151 & 1267 \\
%B & 06 17 29.4  & +22 23 38  &  129(20) &  -6.18(0.08)  &  0.98(0.18)  & 15.2  &  9.4  &  84 & 25 \\
%D & 06 17 57.6  & +22 24 34  &  163(8)  &  -6.97(0.04)  &  1.35(0.08)  & 13.3  &  9.3  &  67 & 23 \\
\hline
\end{tabular}
\label{tbl:gaussfits}
\\
{Errors to fits are shown in parenthesis.}
\end{table}

%table 2, absorption fits
%\begin{table}[h]
%\centering
%\caption{OH Absorption Gaussian Fits}
%\vspace{10pt}
%\begin{tabular}{c|rrr|rrr|rrr|}
%\hline
% & \multicolumn{3}{c}{1667MHz} & \multicolumn{3}{c}{1665MHz} & \multicolumn{3}{c|}{1612MHz} \\
%\cline{2-4} \cline{5-7} \cline{8-10}
%Clump & V$_{LSR}$  & S$_P$ & $\Delta$$V$  & V$_{LSR}$  & S$_P$ & $\Delta$$V$  & V$_{LSR}$  & S$_P$ & $\Delta$$V$ \\
%      & \kms\      & mJy   & \kms\        & \kms\      & mJy   & \kms\        & \kms\      & mJy   & \kms\ \\
%\hline
%G  & --3.06 & --241 &  2.55 &  --2.92 & --155 &  5.31 &  --3.33 & --169 & 4.06 \\
%   & --5.77 & --155 &  2.10 &  --8.89 & --154 & 17.36 &  --6.84 &  --96 & 20.32\\
%B  & --3.21 & --284 &  1.16 &  --2.34 & --205 &  1.04 &         &       &      \\
%   &--15.07 & --559 & 20.99 & --16.06 & --335 &  1.83 & --16.77 & --102 & 29.99\\
%   &--43.39 & --182 & 49.75 & --36.69 & --117 &  2.92 &         &       &      \\
%D  & --4.32 & --101 &  5.74 &         &       &       &         &       &      \\
%   & --9.44 & --147 &  2.78 & --12.07 & --140 & 13.60 &  --7.84 &  --93 & 12.66\\
%   &--14.71 & --146 & 20.6  & --22.75 &  --59 &  3.85 &         &       &\\
%\hline
%\end{tabular}
%\end{table}

%table 3, model parameters
\begin{table}[h]
 \centering
 \caption{Model parameters}\vspace{12pt}
 \begin{tabular}{l|rr|rr|rr|}
     \hline
                           &    B   &   B   &    D   &   D   &    G   &   G   \\
                           & narrow & broad & narrow & broad & narrow & broad \\
     \hline                                                                        
 $v_0(\mathrm{km/s})$      & --6.1  &--14.2 &  --6.8 &--11.0 &  --4.6 & --6.4 \\
 $\Delta v(\mathrm{km/s})$ &   1.8  &    23 &    3.5 &    19 &    1.8 &    16 \\
 $\noh(10^{16}\,                                                                 
 \mathrm{cm}^{-2})$        &   0.7  &   1.1 &    3.1 &   0.7 &   14.0 &   9.8 \\
       $T_k (K)$           &    35  &    50 &     35 &    45 &     60 &    44 \\
      $f$                  &  0.07  &     1 &   0.03 &     1 &   0.05 &   0.3 \\
     \hline                                                                       
 $\tex$(1612) (K)          &   4.0  &   6.8 &    4.6 &   8.3 &    3.1 &   9.3 \\
 $\tex$(1665) (K)          &  18.5  &   7.3 &   19.1 &   8.9 &    7.5 &   9.7 \\
 $\tex$(1667) (K)          &  13.7  &   7.5 &   15.9 &   9.2 &    7.2 &  10.0 \\
 $\tex$(1720) (K)          & --9.0  &   8.1 & --11.0 &   9.8 & --27.1 &  10.5 \\
     \hline
 $\tau$(1612)              &   0.44 &  0.03 &   0.87 &  0.02 &  11.4  &  0.30 \\
 $\tau$(1665)              &   0.49 &  0.15 &   1.09 &  0.10 &  24.3  &  1.48 \\
 $\tau$(1667)              &   1.20 &  0.27 &   2.35 &  0.17 &  45.7  &  2.59 \\
 $\tau$(1720)              & --0.21 &  0.03 & --0.39 &  0.02 & --1.39 &  0.28 \\
     \hline
 \end{tabular}
     \label{tbl:model}
\end{table}

\begin{figure}
\plotone{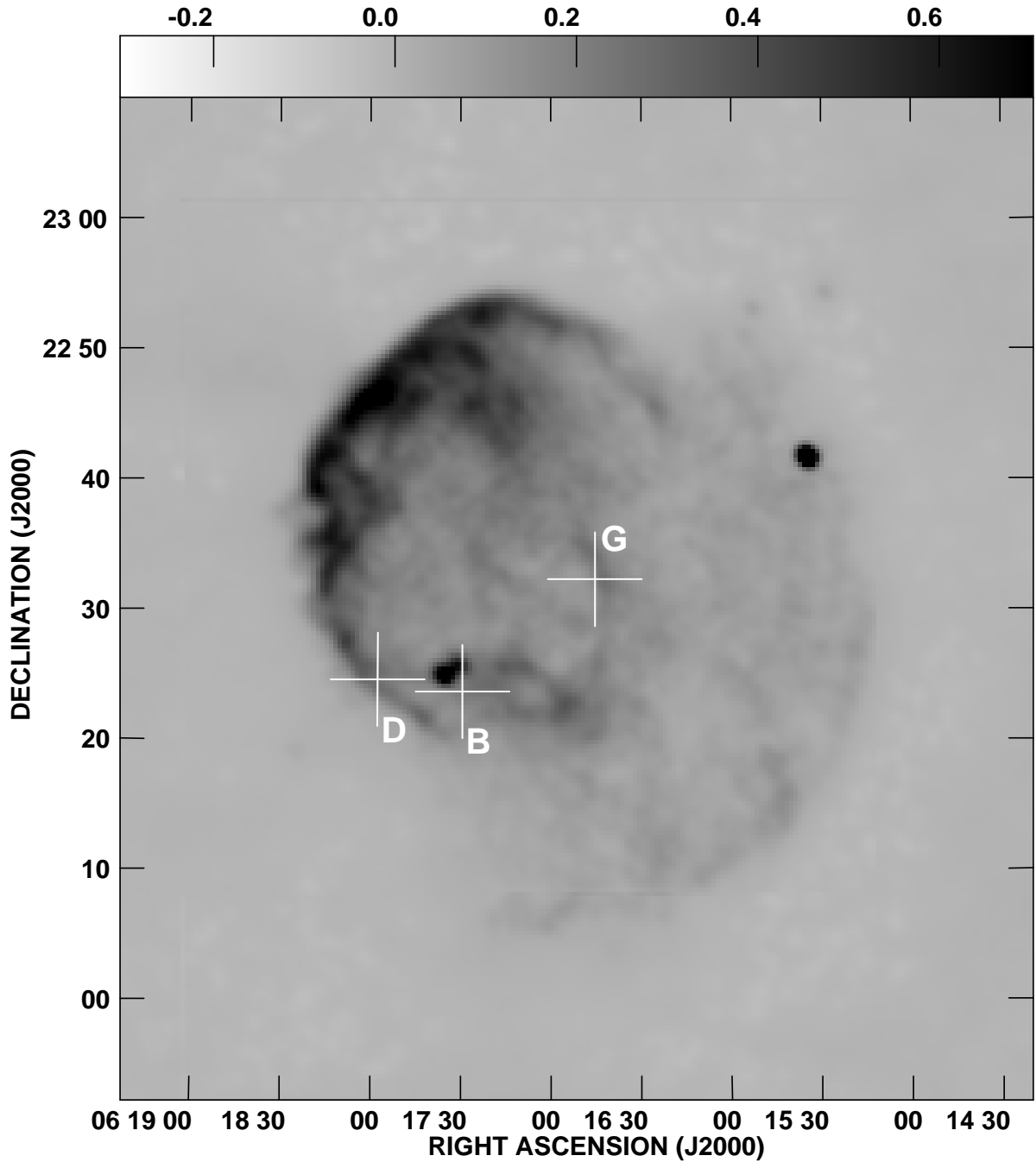}
\caption{Gray-scale continuum image of IC443 at 330 MHz between --300 and 700 mJy beam$^{-1}$ at a resolution of 74$\arcsec\times$64$\arcsec$. Positions of maser emission are demarcated with crosses which extend one GBT beam-width (7$\darcmin$2).}
\end{figure}

\begin{figure}
\plotone{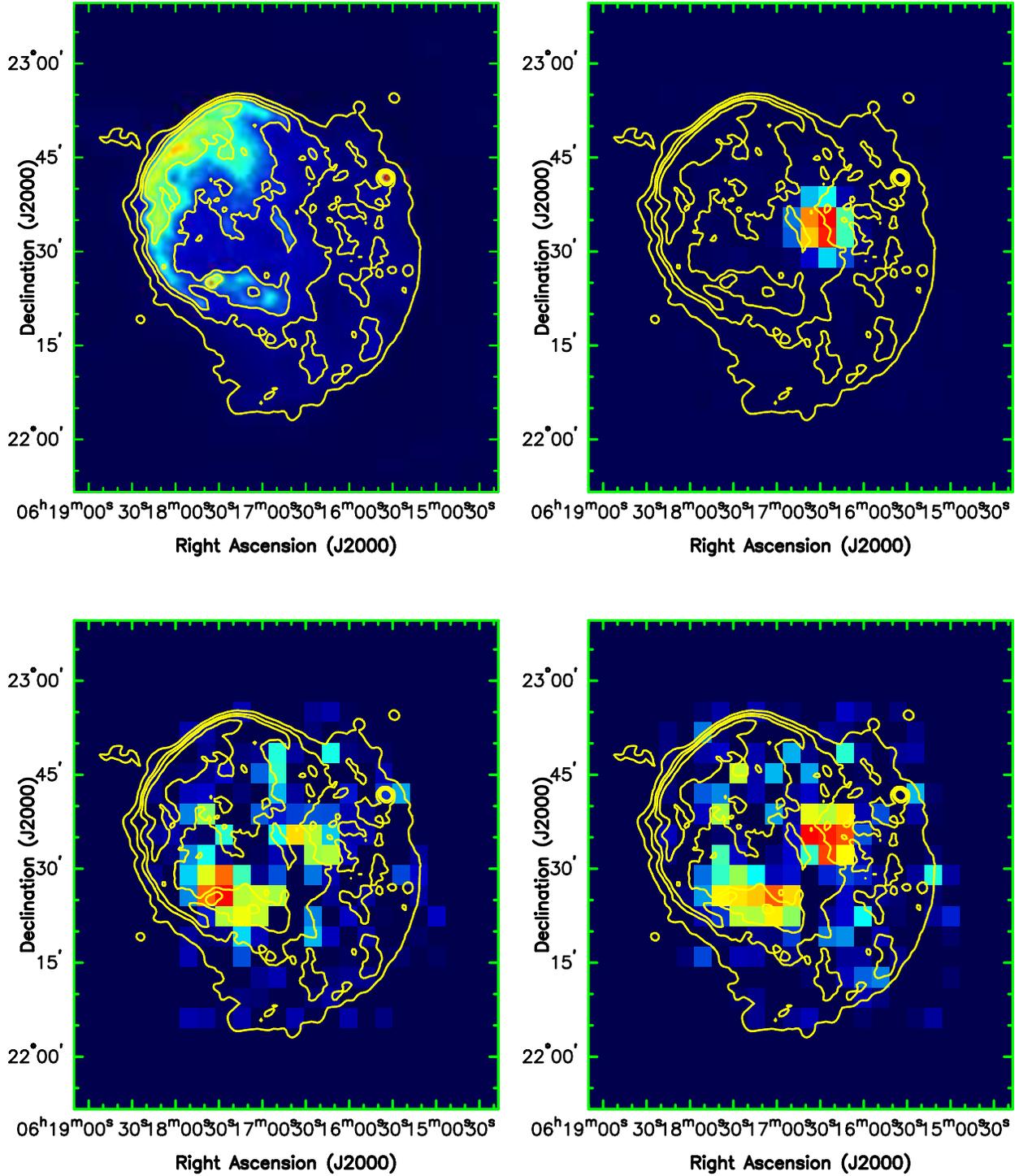}
\label{fig:continuum}
%\centering
%\includegraphics[scale=0.8]{f2.eps}
\caption{Top left: 330MHz continuum map of IC443; Top right: the velocity integrated line flux density (moment zero) map of maser G at -4.55 \kms . Bottom right: moment zero map of maser B at -6.14 \kms ; Bottom left: moment zero map of maser D at -6.85 \kms . Contours of 330 MHz continuum are present for all figures at 0.01, 0.05, 0.1 and 0.3 times the peak continuum level of 177 mJy beam$^{-1}$. Note that due to the brightness of maser G its emission wings are seen in all three moment zero maps.}
\end{figure}

\begin{figure}
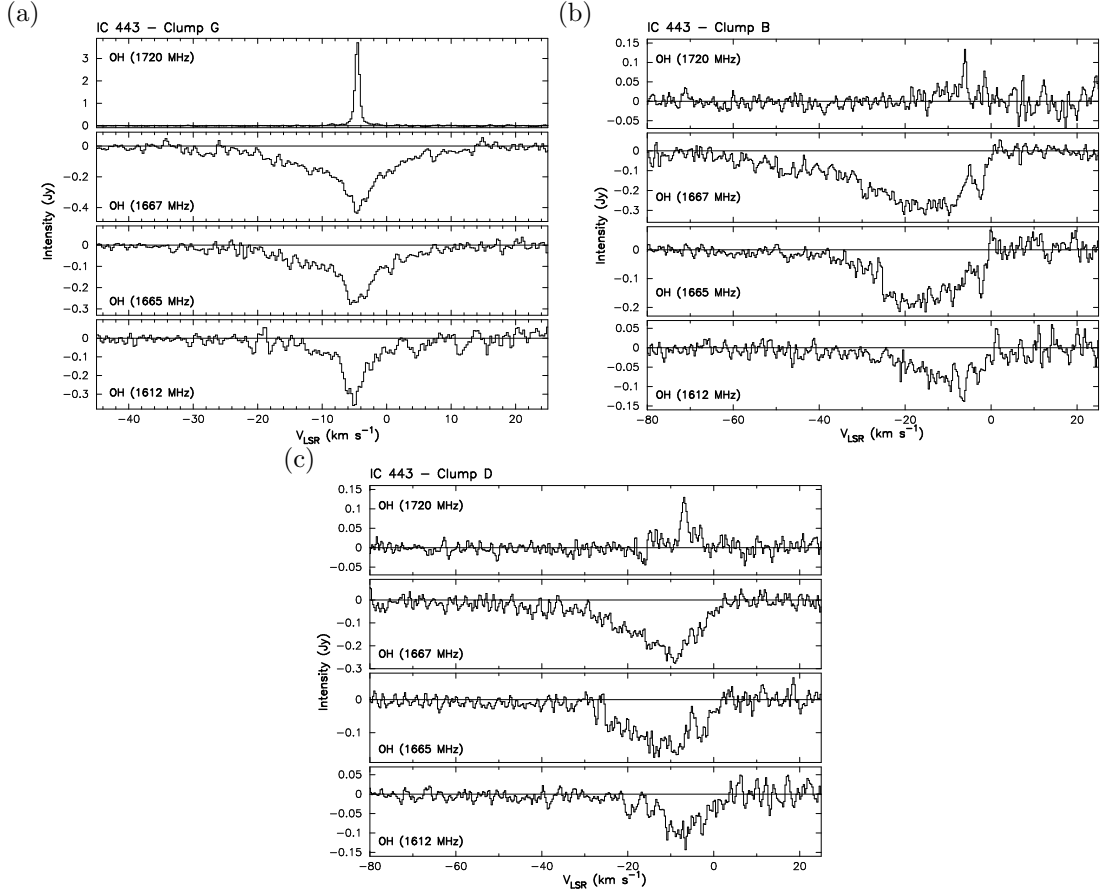

\notetoeditor{Figure 3 contains three eps files -- f3a.eps, f3b.eps, f3c.eps -- which are part of one figure. Please label each with a,b,c.}
\figurenum{3}
\centering
(a)\includegraphics[angle=270,scale=0.3]{f3a.eps}
(b)\includegraphics[angle=270,scale=0.3]{f3b.eps}
(c)\includegraphics[angle=270,scale=0.3]{f3c.eps}
\caption{Spectra of all four ground-state transitions of OH from pointings
taken toward three clumps with detected 1720 MHz emission. 
(a) Clump G at $\alpha$,$\delta$(J2000) = 6$^{\rm h}$16$^{\rm m}$45$\dsec$4,
+22$\degr$32$\arcmin$16$\arcsec$. A strong, narrow emission feature is seen
at 1720 MHz, whereas broad absorption is seen at the other transitions of OH,
indicating a transverse shock.
(b) Clump B at $\alpha,\delta$(J2000) =
6$^{\rm h}$17$^{\rm m}$29$\dsec$4, +22\degr23\arcmin38\arcsec.
A narrow emission feature is seen at 1720 MHz just above the 3$\sigma$ level,
whereas broad absorption is seen at the other transitions of OH, with a sharp
absorption line representing the preshock gas at V$_{LSR}$ = --3.21 \kms.
(c) Clump D at $\alpha,\delta$(J2000) =
6$^{\rm h}$17$^{\rm m}$57$\dsec$6, +22\degr24\arcmin34\arcsec.
A narrow emission feature is seen at 1720 MHz just above the 3$\sigma$ level,
whereas broad absorption is seen at the other transitions of OH, with a sharp
absorption line representing the pre-schock gas at V$_{LSR}$ = --4.32 \kms.}
\end{figure}

\begin{figure}
\notetoeditor{Figure 4 contains three eps files -- f4a.eps, 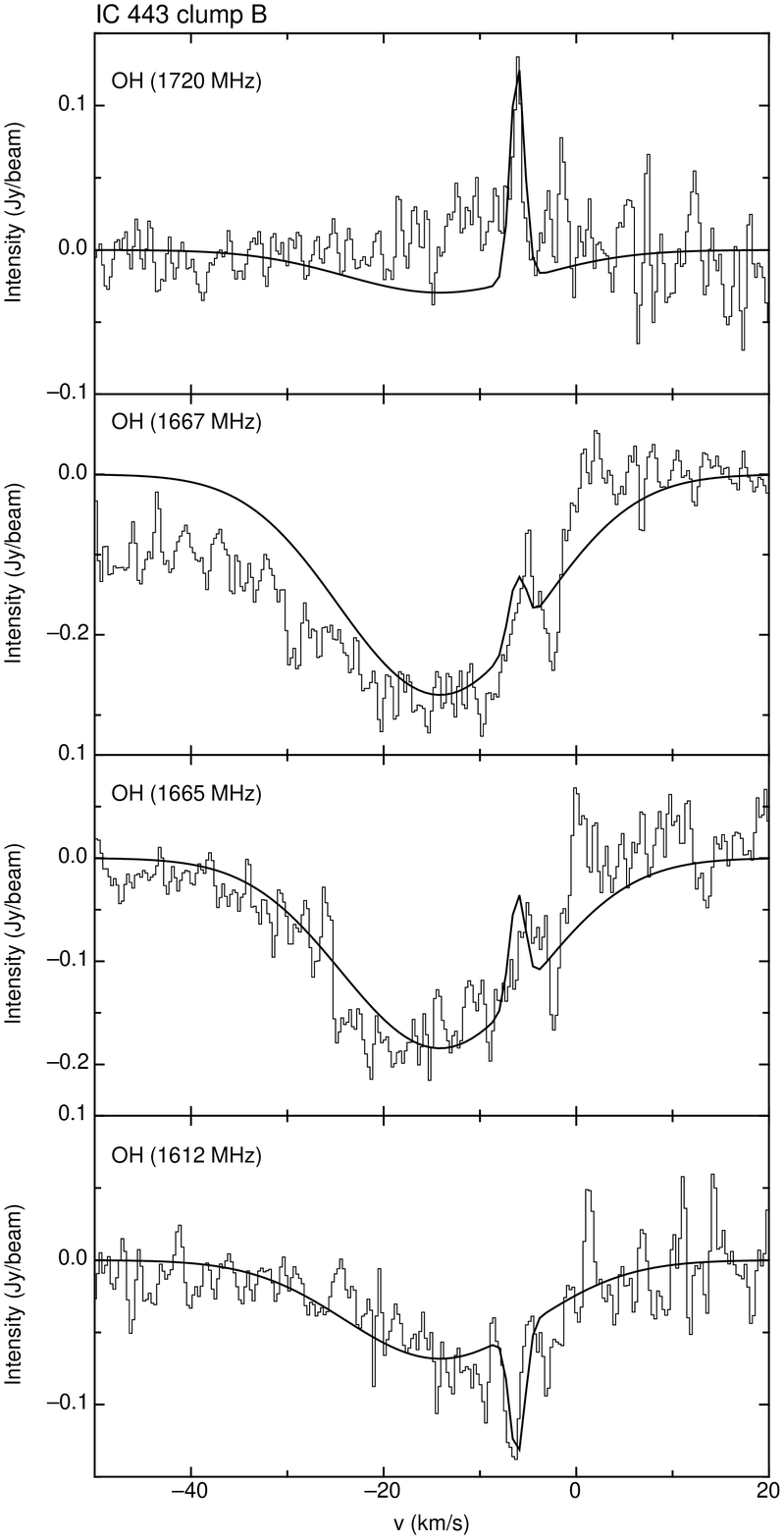, 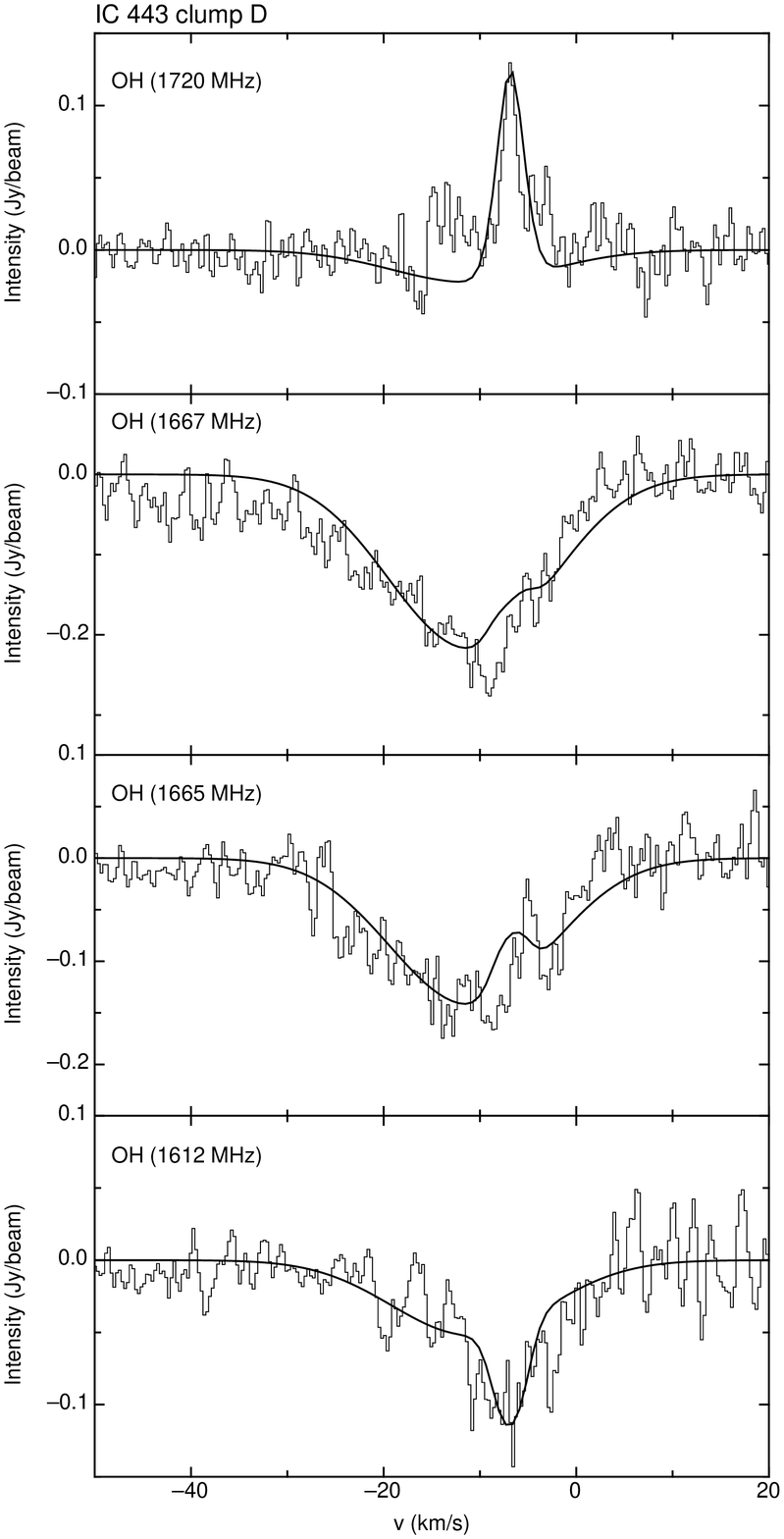 -- which are part of one figure. Please label each with a,b,c.}
\figurenum{4}
\centering
(a)\includegraphics[angle=0,scale=0.3]{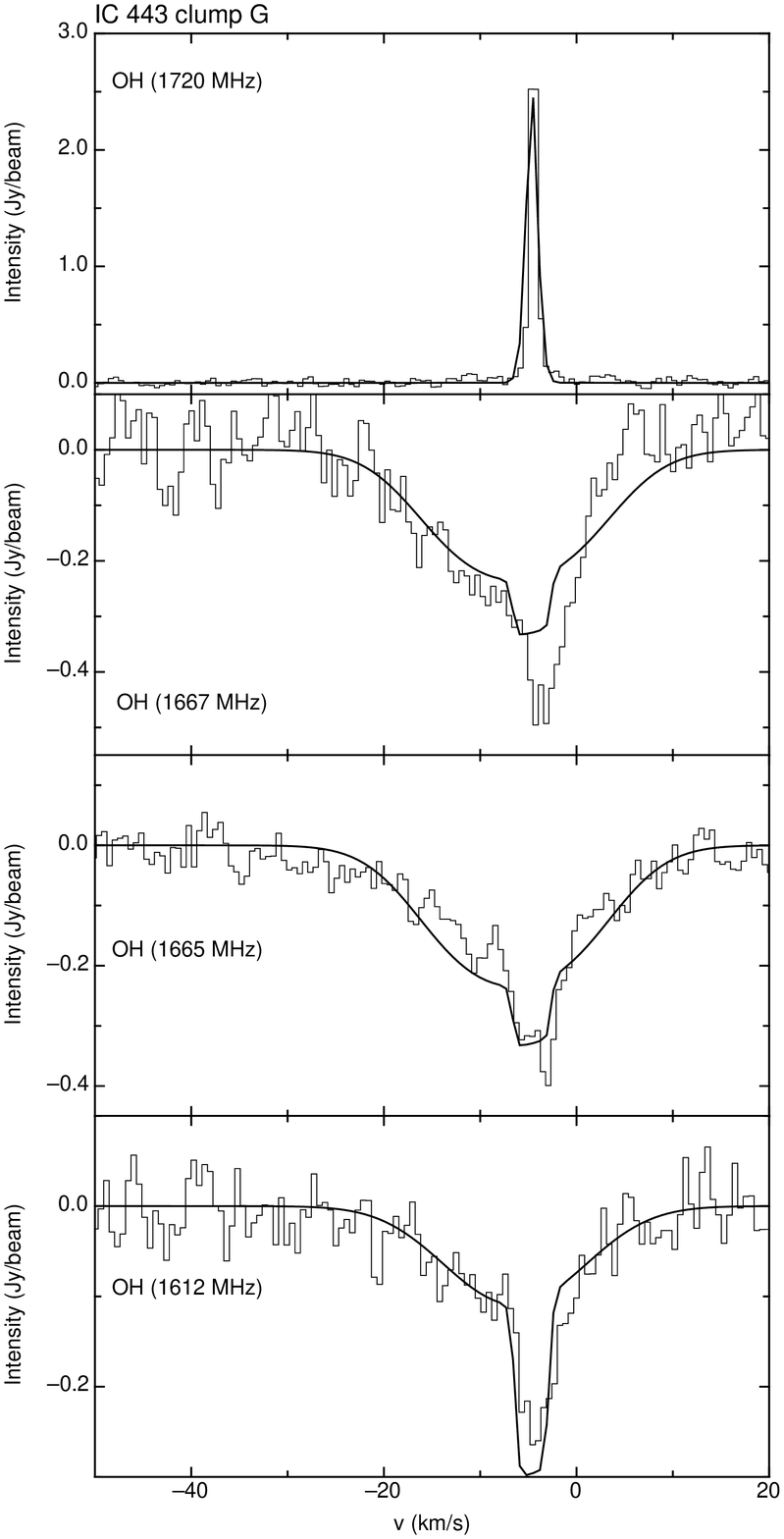}
(b)\includegraphics[angle=0,scale=0.3]{f4b.eps}
(c)\includegraphics[angle=0,scale=0.3]{f4c.eps}
\caption{Spectra of all four ground-state transitions of OH from pointings
taken toward three clumps with detected 1720 MHz emission. The solid line
shows the best fit model for the observed spectrum.
(a) Clump G at $\alpha$,$\delta$(J2000) = 6$^{\rm h}$16$^{\rm m}$45$\dsec$4,
+22$\degr$32$\arcmin$16$\arcsec$. A strong, narrow emission feature is seen
at 1720 MHz, whereas broad absorption is seen at the other transitions of OH,
indicating a transverse shock.
(b) Clump B at $\alpha,\delta$(J2000) =
6$^{\rm h}$17$^{\rm m}$29$\dsec$4, +22\degr23\arcmin38\arcsec.
A narrow emission feature is seen at 1720 MHz just above the 3$\sigma$ level,
whereas broad absorption is seen at the other transitions of OH, with a sharp
absorption line representing the preshock gas at V$_{LSR}$ = --3.21 \kms.
(c) Clump D at $\alpha,\delta$(J2000) =
6$^{\rm h}$17$^{\rm m}$57$\dsec$6, +22\degr24\arcmin34\arcsec.
A narrow emission feature is seen at 1720 MHz just above the 3$\sigma$ level,
whereas broad absorption is seen at the other transitions of OH, with a sharp
absorption line representing the pre-schock gas at V$_{LSR}$ = --4.32 \kms.}
\end{figure}


\begin{thebibliography}{} 

\bibitem[Braun \& Strom 1996]{braun96} Braun, R. \& Strom, R.G. 1996, A\&A, 164, 193

\bibitem[Burton 1987]{burton87} Burton, M. 1987, Q. Jl R. Astr. Soc. 28, 269

\bibitem[Claussen et al. 1997]{claussen97} Claussen, M.J., Frail, D.A., Goss, W.M., \& Gaume, R.A. 1997, ApJ, 489, 143

\bibitem[Cornwell \& Perley 1992]{cornwell} Cornwell, T.J. \& Perley, R.A. 1992, A\&A, 261, 353

\bibitem[Crutcher 1977]{cru77} Crutcher, R.M. 1977, ApJ, 216, 308

\bibitem[DeNoyer 1978]{denoyer78} DeNoyer, L.K. 1978, MNRAS, 183, 187

\bibitem[DeNoyer 1979]{denoyer79} DeNoyer, L.K. 1979, ApJ, 232, L165

\bibitem[DeNoyer \& Frerking 1981]{denoyer1981} DeNoyer, L.K. 1981, ApJ, 246, 37

\bibitem[Dickman et al. 1992]{dickman1992} Dickman, R.L., Snell, R.L., Ziurys, L.M. \& Huang, Y. 1992, ApJ, 400, 203

\bibitem[Erickson \& Mahoney 1985]{erickson} Erickson, W.C. \& Mahoney, M.J. 1985, ApJ, 290, 596.

%\bibitem[Fesen \& Kirshner 1980]{fesen80} Fesen, R.A. \& Kirshner, R.P. 1980, ApJ, 242, 1023

\bibitem[Frail, Goss \& Slysh 1994]{frail94} Frail, D.A., Goss, W.M., \& Slysh, V.I. 1994, ApJ, 424, L111

\bibitem[Frail et al. 1996]{frail96} Frail, D.A., Goss, W.M., Reynoso, E.M., Giacani, E.B., Green, A.J. \& Otrupcek, R. 1996, AJ, 111, 1651

\bibitem[Goss, W.(1968)]{gos1968} Goss, W.M. 1968, ApJS, 15, 131

\bibitem[Hartman 1999]{hartman99} Hartman, R.C. et al.. 1999, ApJS, 123, 79

\bibitem[Hoffman et al. 2003]{hoffman03} Hoffman, I.M., Goss, W.M., Brogan, C.L., Claussen, M.J. \& Richards, A.M.S. 2003, ApJ, 583, 272

\bibitem[Huang, Dickman, \& Snell 1986]{huang86} Huang, Y.L., Dickman, R.L. \& Snell, R.L. 1986, ApJ, 302, L63

\bibitem[Lockett, Gauthier \& Elitzur 1999]{lockett99} Lockett, P., Gauthier, E. and Elitzur, M. 1999 ApJ, 511, 235

%\bibitem[Rho et al. 2001]{rho2001} Rho, J., Jarrett, T.H., Cutri, R.M. and Reach, W.T. 2001, ApJ, 547, 885

\bibitem[Seta et al. 1998]{seta1998} Seta, M., Hasegawa, T., Dame, T.M., Sakamoto, S., Oka, T., Handa, T., Hayashi, M., Morino, J., Sorai, K. \& Usuda, K.S. 1998, ApJ, 505, 286

\bibitem[Snell et al. 2005]{snell2005} Snell, R.L., Hollenbach, D., Howe, J.E., Neufeld, D.A., Kaufman, M.J., Melnick, G.J., Bergin, E.A. and Wang, Z. 2005, ApJ, 620, 758

\bibitem[Tauber et al. 1994]{tauber94} Tauber, J.A., Snell, R.L., Dickman, R.L. and Ziurys, L.M. 1994, ApJ, 421, 570

\bibitem[van Dishoeck, Jansen \& Phillips 1993]{vanDishoeck93} van Dishoeck, E.F, Jansen, D.J. \& Phillips, T.G. 1993, A\&A, 279, 541

%\bibitem[Wang \& Scoville 1992]{wang92} Wang, Z. \& Scoville, N.Z. 1992, ApJ, 386, 158

\bibitem[Wardle 1999]{wardle99} Wardle, M. 1999, ApJ, 525, L101

\bibitem[Yusef-Zadeh, Uchida \& Roberts 1995]{fyz95} Yusef-Zadeh, F., Uchida, K.I. \& Roberts, D.A. 1995, Science, 270, 1801

\bibitem[Yusef-Zadeh et al. 1999]{fyz99} Yusef-Zadeh, F., Goss, W.M., Roberts, D.A., Robinson, B. and Frail, D.A. 1999, ApJ, 527, 172

\end{thebibliography}
\end{document}